\documentclass[final,11pt,3p,sort&compress]{elsarticle} 
\makeatletter
\def\ps@pprintTitle{%
 \let\@oddhead\@empty
 \let\@evenhead\@empty
 \def\@oddfoot{\reset@font\hfil\thepage\hfil}%
 \let\@evenfoot\@oddfoot}

\biboptions{super}	

\usepackage[T1]{fontenc} 
\usepackage[latin2]{inputenc}
\usepackage{amssymb}
\usepackage{amsthm}

\usepackage{graphicx,cancel}
\usepackage{ulem}
\usepackage{color}
\usepackage{fixltx2e} 
\usepackage{dblfloatfix} 
\usepackage{booktabs} 
\usepackage{microtype} 
\usepackage[
   separate-uncertainty  = true,
   mode = math,
   multi-part-units      = brackets,
]{siunitx} 
\usepackage{hyperref}
\hypersetup{
    colorlinks=true, 
    linkcolor=blue,  
    citecolor=blue,  
}
\usepackage[all]{hypcap}  

\renewcommand{\_}[1]{{}_{\mathrm{#1}}}

\begin{document}

\begin{frontmatter}

\title{Photoelectron spectroscopy investigation of the temperature-induced deprotonation and substrate-mediated hydrogen transfer in a hydroxyphenyl-substituted porphyrin}

\author[afko]{Lars~Smykalla\corref{cor1}}
\author[afko]{Pavel~Shukrynau}
\address[afko]{{\small Technische Universität Chemnitz, Institute of Physics, Solid Surfaces Analysis Group, D-09107 Chemnitz, Germany}}
\author[chem]{Carola~Mende}
\author[chem]{Heinrich~Lang}
\address[chem]{{\small Technische Universität Chemnitz, Institute of Chemistry, Inorganic Chemistry, D-09107 Chemnitz, Germany}}
\author[IFW]{Martin~Knupfer}
\address[IFW]{{\small Electronic and Optical Properties Department, IFW Dresden, D-01171 Dresden, Germany}}
\author[afko]{Michael~Hietschold}

\date{\today}

\begin{abstract}

The temperature dependent stepwise deprotonation of 5,10,15,20-tetra\-(\textit{p}-hydroxyphenyl)\-porphyrin is investigated using photoelectron spectroscopy. An abundance of pyrrolic relative to iminic nitrogen and a decrease in the ratio of the amount of $-$NH$-$ to $-$N= with increasing annealing temperature is found. In contrast to the molecules adsorbed on Au(111), on the more reactive Ag(110) surface, partial dissociation of the hydroxyl groups and subsequent diffusion and rebonding of hydrogen to the central nitrogen atoms resulting in a zwitterionic molecule was clearly observed. Moreover, partial C$-$H bond cleavage and the formation of new covalent bonds with adjacent molecules or the surface starts at a relatively high annealing temperature of \SI{300}{\celsius}. This reaction is identified to occur at the carbon atoms of the pyrrole rings, which leads also to a shift in the N 1s signal and changes in the valence band of the molecules.
Our results show that annealing can significantly alter the molecules which were deposited depending on the maximum temperature and the catalytic properties of the specific substrate. The thermal stability should be considered if a molecular monolayer is prepared from a multilayer by desorption, or if annealing is applied to enhance the self-assembly of molecular structures.

\end{abstract}

\end{frontmatter}

\section{Introduction}

Organic materials are expected to be of crucial importance for the construction of nano-devices to address tomorrows challenges in electronics, opto-electronics, photonics, and energy and information storage. In this sense, porphyrin-based molecules have been studied in great detail for many years because of their exceptional versatility. These organic molecules can be easily modified to change the intermolecular or molecule-substrate interaction and their optical, electronic and also magnetic properties.\cite{Buchner2008,Ecija2008, Iancu2006, Qiu2004, Rojas2010, Nowakowski2013} This class of molecules is of outstanding importance in biological systems, in which they represent the active centre of many enzymes. Examples are iron porphyrin in heme and magnesium porphyrin in chlorophyll. Meso-tetra\-(hydroxyphenyl)porphyrin can be used effectively as photosensitizer, e.g. for tumour treatment.\cite{Berenbaum1986, Bonnett1989} Furthermore, the self assembly of metal-free 5,10,15,20-tetra(\textit{p}-hydroxyphenyl)porphyrin in water leads to the formation of nano-scaled, hollow spheres\cite{Lu2009}, which have possible applications, for example, in drug delivery and chemical storage. In nanotechnology this class of molecules is promising for various applications, for example, in non-linear optics~\cite{Ogawa2002, Screen2002}, gas sensors~\cite{Rakow2000}, as catalysts~\cite{Fukuzumi2001} and in photochemical and photovoltaic cells~\cite{Tanaka2007, Mozer2009}.
The porphyrin macro-cycle exhibits a high chemical and thermal stability. Nevertheless, the effect of heating and the limit of this stability should be investigated in more detail, as production steps in the fabrication of devices might damage the molecular films or partially decompose the molecules~\cite{Beggan2012} and, therewith, possibly reduce the performance. Investigations on which reactions occur for such molecules when exposed to high annealing temperatures are quite rare. It is known that at a temperature often between circa \SI{250}{\celsius} to \SI{300}{\celsius} desorption of porphyrin multilayer on metals starts and only the molecules which are in direct contact with the substrate remain\cite{Lukasczyk2007,Chen2010,Bai2010}, i.e. those with a higher adsorption energy due to the strong molecule-substrate interaction. This presents a convenient way to produce a surface fully covered with an only one monolayer thick film. It is also known that annealing increases the mobility of the molecules, which aids in or induces the formation of large highly ordered domains in the molecular film.\cite{THPPstruc} By this, the transport properties of devices could be improved. The desorption temperature depends on the functional groups and the coordinated metal ion of the porphyrin core, as well as the substrate. For heating at high temperatures, a change in the shape of the C 1s core level spectra was often observed, which was usually explained by the term ``decomposition'' without further details.\cite{Lukasczyk2007, Chen2010} Herein, a closer look at what the nature of this decomposition really is will be taken to gain knowledge about which bond cleavage reactions occur for porphyrins at high annealing temperature.

Previously, the coexistence of two stable forms of a tetra\-(\textit{p}-hydroxy\-phenyl)\-porphyrin molecule was found by scanning tunneling microscopy. Switching between them was suggested to be due to a local deprotonation or protonation reaction which changes the number of pyrrolic to iminic nitrogen.\cite{THPPspec} Thereby, a significant change in the electronic structure near $E\_F$ was measured. To investigate also this matter further, photoelectron spectroscopy, which can give additional information about the bonding conditions in the molecule, was applied.

\section{Experimental details}

5,10,15,20-tetra(\textit{p}-hydroxy\-phenyl)\-porphyrin (H$\_2$THPP) was synthesized as described in an earlier publication\cite{THPPspec}. Single crystals of Au(111) and Ag(110) were cleaned in ultra-high vacuum (UHV) by cycles of Ar$^+$-sputtering at an energy of \SI{500}{eV} and annealing to \SI{500}{\celsius} for \SI{30}{\minute}. The powder of H$\_2$THPP was purified by heating to a temperature slightly below the sublimation temperature in UHV for several days. A multilayer of the molecules were then deposited on Au(111) (layer thickness around \SI{2.4}{nm}) and on Ag(110) ($\approx \SI{1.8}{nm}$) with organic molecular beam deposition at $\approx \SI{350}{\celsius}$. The temperature of the substrate during deposition was held at room temperature. In the following, these initial layers will be denominated as ``thick layer''. The thicknesses of the layers were estimated with a quartz micro-balance and from the attenuation of the respective substrate core-level peaks (Au 4f or Ag 3d). A monolayer (ML) is defined as the minimal thickness where the substrate surface is completely covered with close-packed molecules ($\approx \SI{0.35}{nm}$). Photoelectron spectroscopy (PES) experiments were performed with synchrotron radiation at the Material Science Beamline in Elettra (Trieste, Italy). A Phoibos photoelectron spectrometer was used. All spectra were measured in a geometry for emission along the surface normal (\SI{30}{\degree} incidence of beam relative to the surface plane). The excitation photon energies used for acquiring the C 1s, N 1s and O 1s core level spectra of H$\_2$THPP were for Ag(110)/Au(111) \SI{390}{eV}/\SI{400}{eV}, \SI{500}{eV}/\SI{520}{eV} and \SI{650}{eV}, respectively, and \SI{22}{eV} for the valence band spectra.
The binding energies were corrected relative to the Au 4f$\_{7/2}$ peak at \SI{83.98}{eV} or Ag 3d$\_{5/2}$ at \SI{368.21}{eV}. Core-level spectra were fitted with a Shirley background and Voigt peak functions using an identical width for all peaks. The full width at half maximum for example for N 1s ($E\_{exc} = \SI{520}{eV}$, $E\_{pass} = \SI{10}{eV}$) was circa \SI{1.1}{eV}. The typical precision for the energy position of each peak component was $\pm \SI{0.05}{eV}$. Annealing of the sample was performed by resistive heating of a wire wound around the crystal with the thermocouple located at its backside. In each annealing step, the temperature was increased slowly till the designated maximum temperature, where it was held at for \SI{10}{\minute}, and then cooled down to room temperature for the measurements.

\section{Results and Discussion}

\subsection{Protonation and deprotonation at nitrogen atoms}

\begin{figure}[tb]
\centering
\includegraphics[width=0.85\textwidth]{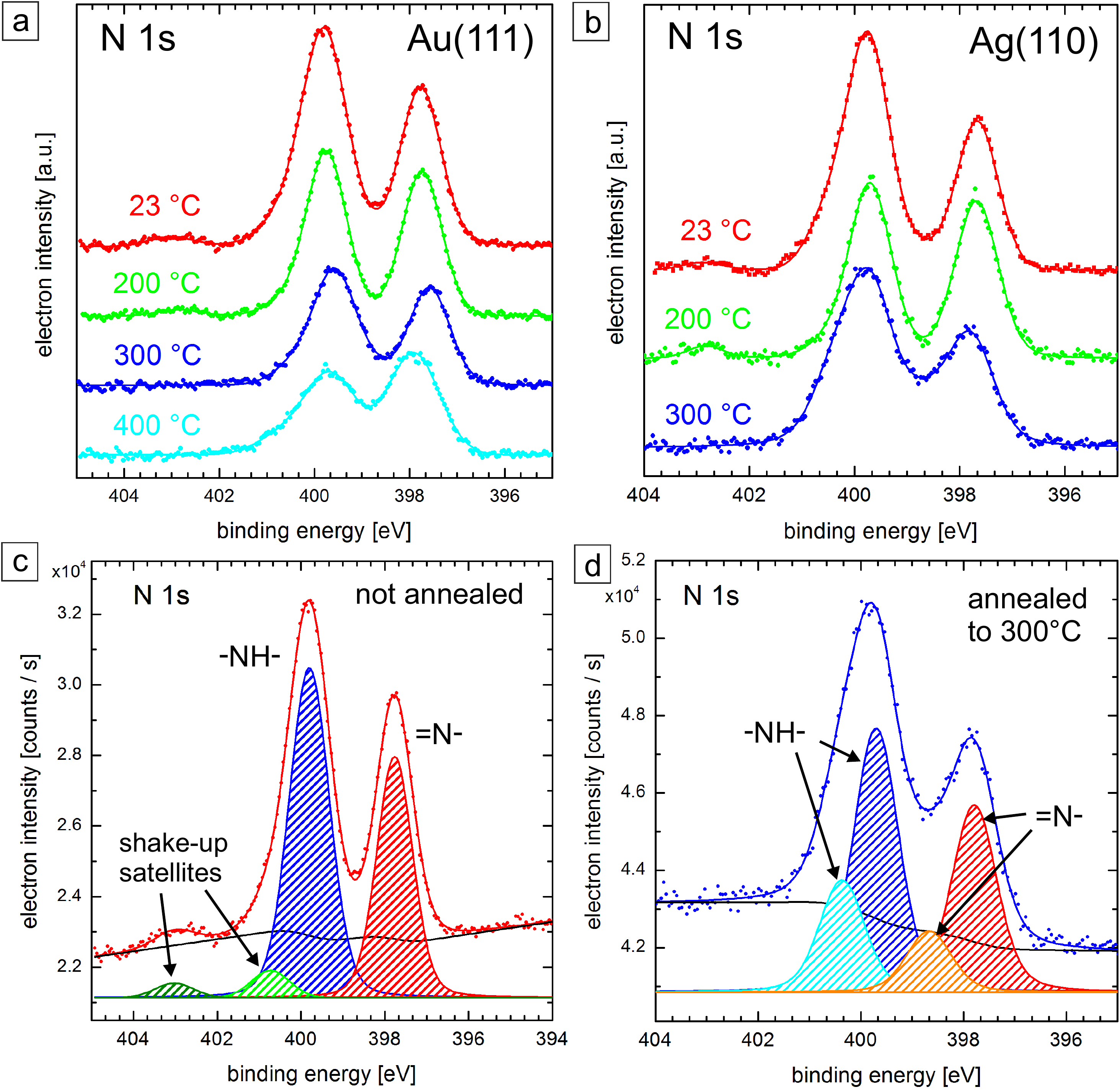}
\caption{Evolution of the PES N 1s core level spectra of an initially thick layer of H$\_2$THPP on Au(111) (\textbf{a}) and on Ag(110) (\textbf{b}) for annealing steps of increasing temperature. Dots corresponds to the measured data points, solid lines to the fit of the signal. The peaks are shifted vertically for better visualization. Fitted N 1s spectra of H$\_2$THPP: (\textbf{c}) thick molecular layer prior to annealing and (\textbf{d}) the layer on Ag(110) after annealing to \SI{300}{\celsius}.}
\label{fig:PES_N1s_Au_Ag}
\end{figure}

The evolution with annealing of the N 1s spectra, which are sensitive to the nitrogen bonded hydrogen atoms of the molecule, is shown in Fig.~\hyperref[fig:PES_N1s_Au_Ag]{\ref*{fig:PES_N1s_Au_Ag}(a),(b)} for H$\_2$THPP on Au(111) and Ag(110), respectively. The assignment of the two-component signal is done in agreement with studies on tetraphenylporphyrin\cite{Buchner2008, Nardi2010, Diller2012, Goldoni2012} with the peak at \SI{399.8}{eV} corresponding to pyrrolic nitrogen, which has a bond to hydrogen ($-$NH$-$), and the right peak at \SI{397.8}{eV} to iminic nitrogen ($-$N=). It can be seen that in the molecular films as deposited [Fig.~\hyperref[fig:PES_N1s_Au_Ag]{\ref*{fig:PES_N1s_Au_Ag}(c)}], the amount of pyrrolic N was clearly larger than iminic N with very similar values on both Au(111) and Ag(110) (ratio 3:2). The expected ratio of both nitrogen species from the chemical structure of H$\_2$THPP is 1:1. Thus, both H$\_2$THPP (80\%) and a protonated form H$\_4$THPP (20\%) were likely present in the molecular layer. A similar phenomenon of non-equivalent N 1s peaks was reported for other porphyrin molecules several times before with different or no explanations given.\cite{Garcia-Lekue2012, Gonzalez-Moreno2011, Goldoni2012, Niwa1974}
Diller \textit{et al}.\cite{Diller2012} found a ratio of the peak areas of the pyrrolic divided by the iminic nitrogen of 2.5 for a monolayer of H$_2$TPP on Cu(111) at normal emission geometry and higher ratios at other angles of emission, whereas for the multilayer the ratio was nearly 1. This behavior was explained by a photoelectron diffraction effect and a strong nitrogen-surface interaction.\cite{Diller2012} To investigate the influence of the geometry for H$\_2$THPP, measurement were done at angles of emission of \SI{90}{\degree}, \SI{70}{\degree} and \SI{30}{\degree} - no change in the ratio of pyrrolic to iminic nitrogen ($\approx 1.6$) was observed. From the different behavior with coverage and no dependence on the angle of emission in our case, photoelectron diffraction effects can be ruled out as a possible reason.
It is known that some porphyrin derivatives, e.g. with sulfonatophenyl~\cite{Friesen2011} or carboxyphenyl~\cite{Yoshimoto2008} groups, can exist as stable neutral zwitter\-ionic (or ``diacid'') molecules. In solutions with low pH values not only acidic groups of the molecule are partially protonated but the iminic nitrogens are protonated as well due to close pK$\_a$ values, thereby creating a positive charge in the central region and a negative charge on the periphery of the molecule. Typically, molecules with such functional groups are deposited from an acidic solution on the substrate and can then be transferred into UHV for measurements.\cite{Friesen2011} Although in our experiments, molecular films were prepared by organic molecular beam deposition, a similar hydrogenation effect was found for metal-free tetra(hydroxyphenyl)porphyrin. The additional hydrogen atoms necessary to form H$\_4$THPP could originate from H$\_2$ which is always present in vacuum or when the molecular powder was exposed to air.


The N 1s spectra measured after annealing to \SI{300}{\celsius} changed their shape and are fitted the best with two peaks additional to the already discussed ones [Fig.~\hyperref[fig:PES_N1s_Au_Ag]{\ref*{fig:PES_N1s_Au_Ag}(d)}]. These newly occurring peaks are located at a binding energy of \SI{400.4}{eV} and \SI{398.7}{eV} and show an intensity ratio similar to the pair of main peaks. It should be noted that these cannot be shake-up satellites because of their high relative intensity. Therefore, we assign them also to pyrrolic and iminic nitrogen, respectively, although they have a difference in the surrounding molecular frame which leads to a different partial charge of these N atoms or changed screening of the core-hole and results in a shift to higher binding energy of both N 1s peaks. This will be discussed further in the second part of this study.

\begin{figure}[tb]
\centering
\includegraphics[width=0.85\textwidth]{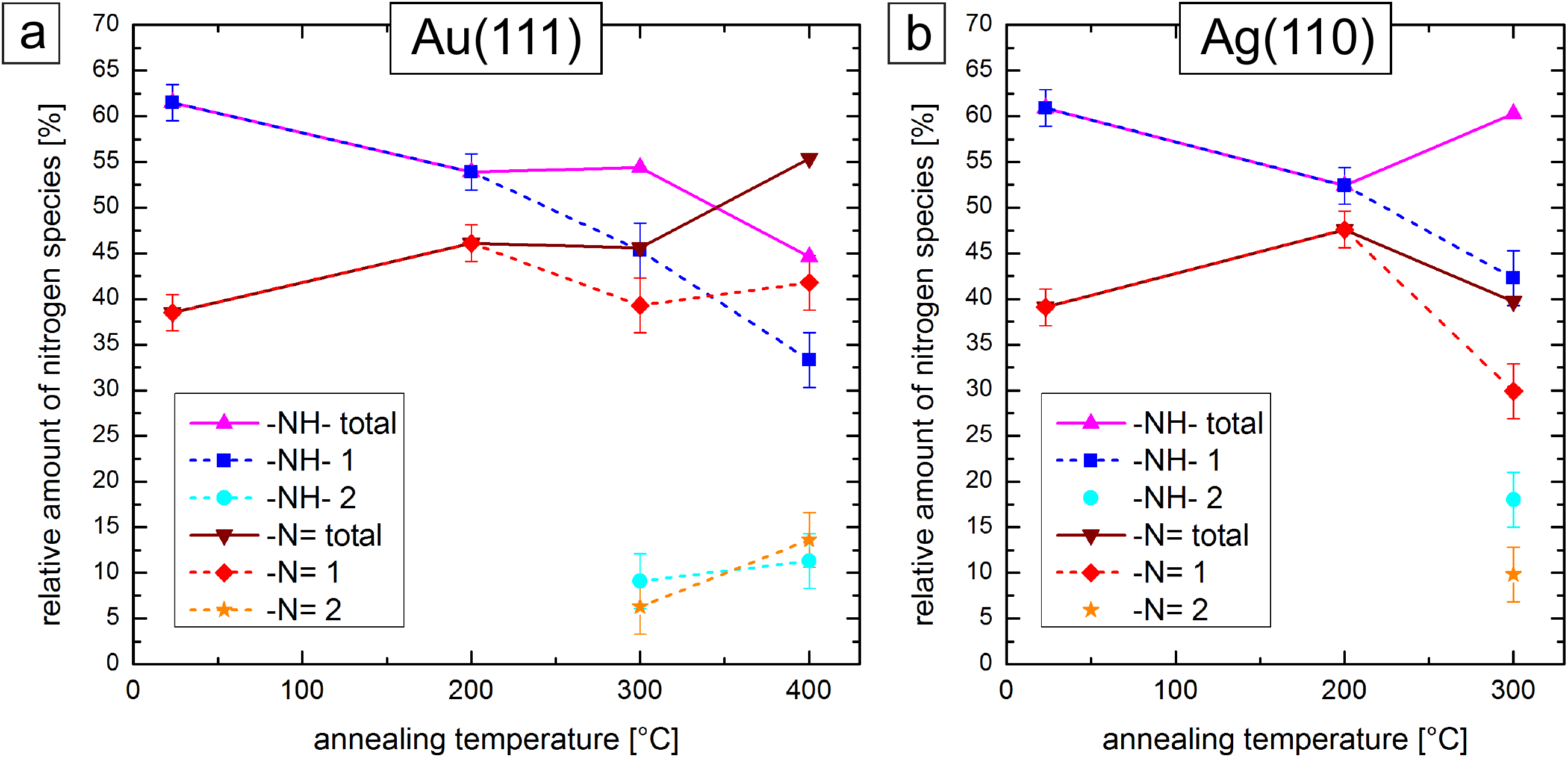}
\caption{Evolution of the relative peak areas of the components in N 1s for H$\_2$THPP on Au(111) (\textbf{a}) and on Ag(110) (\textbf{b}) with increasing temperature of annealing. Lines between the data points are only a guide for the eye.}
\label{fig:N1s_temp}
\end{figure}

The evolution of the relative amount of the N species with annealing is summarized in Fig.~\ref{fig:N1s_temp}. It can be seen that after heating to \SI{200}{\celsius} on both studied substrates the ratio of pyrrolic to iminic N decreased towards 1. This means that the two additional hydrogen atoms in H$\_4$THPP are removed due to a smaller dissociation energy compared to the other two H atoms bonded to N in H$\_2$THPP.
On Au(111) after heating to \SI{300}{\celsius}, the amount of respective total $-$NH$-$ and $-$N= stayed constant. Whereas, after a high maximum temperature (\SI{400}{\celsius}), the total peak intensity of $-$NH$-$ decreased further and became lower than the one for $-$N=. This means that in H$\_2$THPP, the two hydrogen atoms started to dissociate from the nitrogen~\cite{Palmgren2006} and the resulting deprotonated molecule (THPP) is stabilised on the surface.
However for Ag(110), if the sample was heated to \SI{300}{\celsius}, surprisingly, the total amount of $-$NH$-$ [magenta in Fig.~\hyperref[fig:N1s_temp]{\ref*{fig:N1s_temp}(b)}] increased and again a clear inequality of both N species ($-$NH$- > -$N=) was observed for H$\_2$THPP on Ag(110).

\begin{figure}[tb]
\centering
\includegraphics[width=0.97\textwidth]{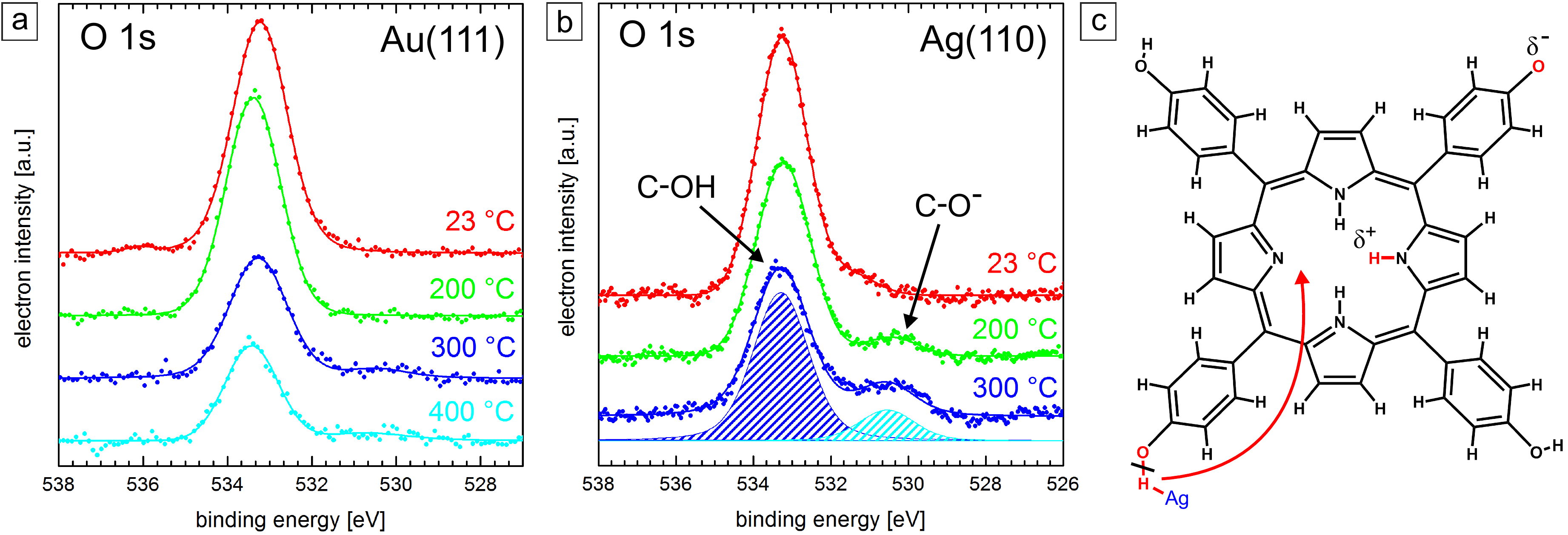}
\caption{Evolution of the O 1s core level spectra with increasing annealing temperature: (\textbf{a}) H$\_2$THPP on Au(111) (\textbf{b}) H$\_2$THPP on Ag(110). The peaks are shifted vertically for better visualization. (\textbf{c}) O$-$H bond cleavage after heating of H$\_2$THPP on Ag(110), and diffusion and bonding of hydrogen from -OH to N.}
\label{fig:PES_O1s}
\end{figure}

To explain the different evolution of the N 1s core level peaks between the adsorption on Au(111) and Ag(110), a look at the oxygen 1s core level spectra (Fig.~\ref{fig:PES_O1s}) is commended. The O 1s peak is located at a binding energy of \SI{533.2}{eV}, which corresponds to the C$-$OH (phenol) environment, i.e. hydroxyl groups. On Au(111), the peak shape does not change significantly with annealing. On the other hand, on Ag(110) after heating to \SI{200}{\celsius}, clearly a new peak at \SI{530.5}{eV} appears. The peak increases with further annealing and after heating to \SI{300}{\celsius} a ratio of $\approx$ 0.21 between both O 1s peaks was measured. A shift to lower binding energy means that this chemical state of oxygen is charged more negative than C$-$OH. Thus, the second peak can be attributed to C$-$O$^-$ (phenolate)\cite{Fischer2014, Chehimi1988}, which means that under influence of the Ag(110) surface the O-H bond was cleaved.\cite{Payer2007,Pawlak2009} The relative number of dissociated H atoms per molecule can be calculated from the ratio of the peak intensities. The amount of H atoms needed for the observed increase of the pyrrolic N 1s peak is, thereby, only slightly lower compared to the dissociated hydrogen as calculated from the O 1s spectra. Thus, it can be concluded that hydrogen atoms from the cleaved hydroxyl groups do not desorb but diffuse on the Ag(110) surface until they rebind to a H$\_2$THPP molecule at the iminic N. This suggested substrate mediated hydrogen transfer results in the formation of a zwitterionic species as shown schematically in Fig.~\hyperref[fig:PES_O1s]{\ref*{fig:PES_O1s}(c)}. It should be noted that, because this process depends on the substrate, only molecules at the molecule-substrate interface should be involved.

A similar effect was reported recently by González-Moreno \textit{et al.}\cite{Gonzalez-Moreno2011} for protoporphyrin molecules on the highly reactive surface of Cu(110), but only at low temperatures because at room temperature a metalation reaction with atoms from the Cu surface occurred. Also, carboxylic groups were found to partially deprotonate even at low temperature, but the amount of possibly migrated H from carboxylic groups to protonated nitrogen was deemed to be not enough to justify the high $-$NH$-$ signal. Thus, it could not be related solely to the formation of the zwitterionic phase. Alternatively, the authors proposed an explanation with hydrogen bonding to the central nitrogen. The crystal structure of H$\_2$THPP~\cite{Hill2007} shows also an arrangement where hydroxyl-groups point towards the central (iminic) nitrogen and likely form $-$N$\cdots$H$\cdots$O$-$ hydrogen bonds. This could also lead to an increase of the $-$NH$-$ and respective decrease of the $-$N= peak~\cite{Garcia-Lekue2012}, and should additionally lead to changes in the O 1s signal. Nevertheless, this alternative explanation would be valid only for quite thick molecular layers, where crystallites with the bulk structure start to grow. In the studied system the adsorption of molecules is parallel to the surface plane for monolayer and few monolayer thick coverages and the molecular arrangements often differ from bulk crystal structures.\cite{THPPstruc} Vertical hydrogen bonding to N in very thick layers would also be independent of the substrate, i.e. this effect would occur for the adsorption on Au(111) as well as Ag(110), which is not the case. Furthermore, after annealing to a maximum of \SI{200}{\celsius}, the N 1s peak ratio is reduced, although heating of molecular films is known from STM investigations to vastly increase the size of ordered areas, which promotes the formation of further hydrogen bonds.\cite{THPPstruc} Thus, an explanation based on hydrogen bonding to the nitrogen atoms seems improbable in our case.

\subsection{Reaction at carbon atoms}

\begin{figure*}[tb]
\centering
\includegraphics[width=0.97\textwidth]{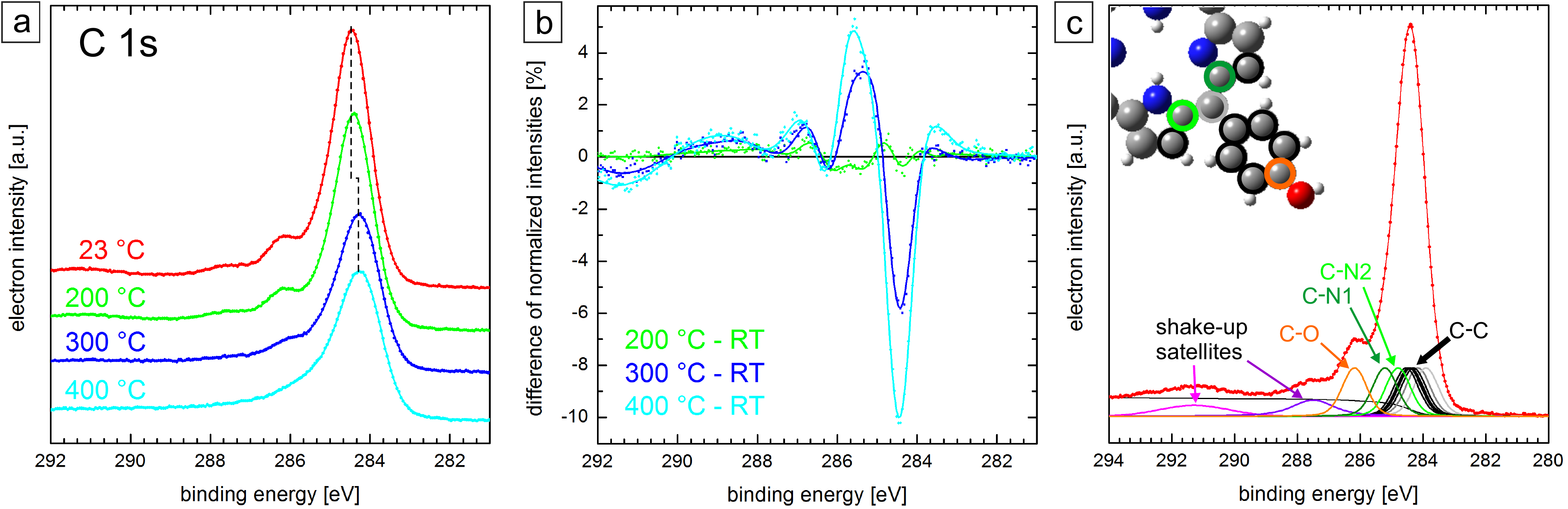}
\caption{(\textbf{a}) Evolution of the C 1s core level spectra of H$\_2$THPP on Au(111) for annealing steps of increasing temperature. The spectra are shifted vertically for clarity. (\textbf{b}) Difference spectra of the annealed layers subtracted with the as deposited layer of H$\_2$THPP to visualize the change of the peak shape. The C 1s signals were first normalized to the total area and shifted to the same energy of the maximum. (\textbf{c}) Deconvolution of the C 1s signal of the not annealed layer and assignment to the 11 inequivalent carbon atoms and shake-up peaks.}
\label{fig:PES_C1s_Au111}
\end{figure*}

In order to understand the appearance of the two new components in the N 1s spectrum after heating to high temperatures [Fig.~\hyperref[fig:PES_N1s_Au_Ag]{\ref*{fig:PES_N1s_Au_Ag}(d)}], one should take a look at the C 1s core level spectra. Fig.~\hyperref[fig:PES_C1s_Au111]{\ref*{fig:PES_C1s_Au111}(a)} shows the evolution of the C 1s spectra for the stepwise annealing of the layer of H$\_2$THPP on Au(111). A very similar change of the shape of the C 1s signal with annealing was found also for H$\_2$THPP on Ag(110) (supplemental information, Fig. S2). The deconvolution of the C 1s spectra is complicated and can be done in several ways. We assumed equal contributions from eleven inequivalent carbon atoms due to the symmetry of the saddle-shaped molecule, whereby each is fitted with a Voigt peak function of identical width and height as well as two smaller, broad peaks for shake-up satellites [Fig.~\hyperref[fig:PES_C1s_Au111]{\ref*{fig:PES_C1s_Au111}(c)}].
By comparison with the C 1s signal of porphyrin molecules without oxygen atoms where a peak at \SI{286.2}{eV} is not observed,\cite{Niwa1974,Chen2010} this component is attributed to C$-$O. The peaks located at \SI{285.2}{eV} and \SI{284.8}{eV} are assigned to the two species of carbon atoms directly bonded to the non-equivalent N atoms (C$-$N) in accordance with theoretical calculations for C 1s of porphine.\cite{Diller2014} Finally, the peaks from \SI{284.5}{eV} to \SI{283.9}{eV} can be grouped into the C$-$C (C$-$H) bonded species.
The broad features centered at \SI{287.5}{eV} and \SI{291.3}{eV} for H$\_2$THPP correspond to superpositions of shake-up satellites similar to H$\_2$TPP.\cite{Lukasczyk2007,Chen2010}
The first $\pi \rightarrow \pi^*$ shake-up peak from the additional HOMO$\rightarrow$LUMO excitation is separated by around \SI{3.0}{eV} from the corresponding main contribution of the C$-$C carbon atoms. For comparison in the N 1s core-level spectra in Fig.~\ref{fig:PES_N1s_Au_Ag}c, the small features from shake-up excitations are also separated by $\approx \SI{3.1}{eV}$ from the corresponding main peaks.

Fig.~\hyperref[fig:PES_C1s_Au111]{\ref*{fig:PES_C1s_Au111}(b)} shows the change of the C 1 peak shape after annealing the H$\_2$THPP/Au(111) sample. For the difference spectra, each C 1s spectrum was normalized with the area of the signal and shifted so that the maximum is at the same energy. Annealing the H$\_2$THPP/Au(111) sample to \SI{200}{\celsius} leads to no significant change of the C 1s spectrum. However, heating to \SI{300}{\celsius} or higher resulted in a broadening of the signal with a decrease of the main contribution at \SI{284.45}{eV} and an increase especially at around \SI{285.5}{eV}.
It should be noted that the relative amount of C to N to O in the molecular layer does not vary significantly with annealing, and that the change in C 1s is not caused by a modification of the nitrogen or oxygen atoms. Thus, the reason is that the number of C$-$H bonded carbon atoms is reduced by thermally activated deprotonation.\cite{Roeckert2014}
If the previously discussed two new shifted peak components in the N 1s signal are a direct result of this, it means that the reaction occurs at the C atoms of the pyrrole rings. The screening of the N atoms could be affected by charge transfer \textit{via} bonding of the deprotonated pyrrolic C atoms to the metal surface or by C$-$C coupling to adjacent molecules.\cite{Zhong2011}
Another possibility might be intramolecular C$-$C bond formation between the carbon atoms of the pyrrole ring and those at ortho-positions of the phenyl group.\cite{Xiao2012} An indication for this reaction and the consequent flattening of H$\_2$THPP was so far not seen by scanning tunneling microscopy.\cite{THPPstruc} However, after high temperature annealing, destruction of ordered adsorbate structures of H$\_2$THPP molecules and the formation of dendrite-like chains was found (supplemental information, Fig. S1), which is similar to the temperature-induced oligomerization observed for phthalocyanines\cite{Manandhar2007}, porphines\cite{Wiengarten2014}, and tetra(mesityl)porphyrin\cite{Veld2008}.
The hydrogen atoms dissociated thereby could then also diffuse and attach at the N atoms as previously discussed. However, compared to a more reactive surface, molecular hydrogen can readily desorb from Au(111) at room temperature under UHV conditions. Thus, the N$-$H peak in N 1s did not increase again on Au(111) even after C$-$H cleavage.

\subsection{Evolution of the valence band}

\begin{figure}[!b]
\centering
\includegraphics[width=0.85\textwidth]{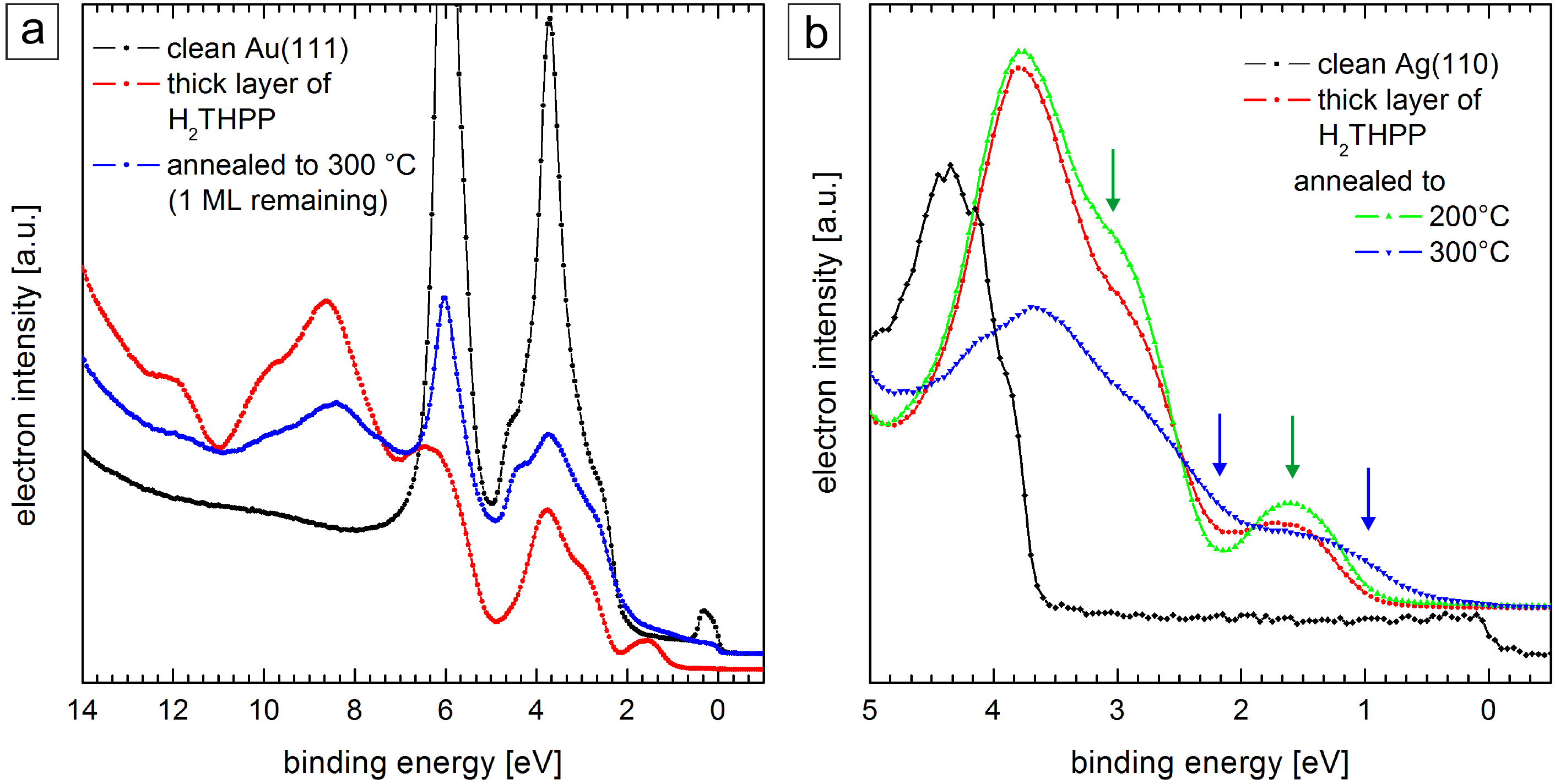}
\caption{(\textbf{a}) Valence band spectra of a monolayer and a thick layer of H$\_2$THPP on Au(111) (\textbf{b}) Valence band near $E\_F$ for H$\_2$THPP on Ag(110) before and after annealing ($E\_{exc}=\SI{22}{eV}$).}
\label{fig:VB}
\end{figure}

The influence of annealing on the valence band was also investigated. Changes due to the coverage are demonstrated for H$\_2$THPP on Au(111) in Fig.~\hyperref[fig:VB]{\ref*{fig:VB}(a)}.
After heating to \SI{300}{\celsius}, multilayers on Au(111) are desorbed with only one monolayer remaining (calculated from the attenuation of Au 4f). This is due to the stronger binding energy of the molecule-substrate interaction compared to intermolecular $\pi$-$\pi$ interactions of stacked molecules. 
Correlated with the desorption of the multilayers is also a shift by about \SI{0.2}{eV} to lower binding energy of all peaks related to the molecule, which can be observed, for example, in Fig.~\hyperref[fig:PES_C1s_Au111]{\ref*{fig:PES_C1s_Au111}(a)} for C 1s of H$\_2$THPP on Au(111). This shift is likely due to the more efficient electrostatic screening of the final hole state in a monolayer by the metal surface as compared to an environment of porphyrin molecules inside the multilayer.\cite{Ishii1999} It can be seen that, when the thick layer is compared to the molecular monolayer (after desorption of multilayer at \SI{300}{\celsius}), all molecular features in the valence band are reduced significantly with decreasing coverage and peaks belonging to the substrate increase accordingly in intensity. Furthermore, at monolayer coverage the Fermi edge of the substrate was again visible but the Shockley-type surface state of Au(111) at $\approx \SI{0.3}{eV}$ below $E\_F$ completely disappeared, which could indicate a small charge transfer to the physisorbed molecules, thereby emptying the surface state.\cite{Rojas2012,Ziroff2009,Xiao2012}

In the case of H$\_2$THPP on Ag(110) only partial multilayer desorption occurred after annealing to \SI{300}{\celsius} with a layer of circa \SI{1}{nm} thickness ($\approx$ 3 ML) remaining. Correspondingly, the $\SI{-0.2}{eV}$ multilayer$\rightarrow$monolayer shift was not observed in the core-level spectra. This indicates also that a complete desorption of multilayer is not an abrupt process, and the necessary temperature (activation energy) is slightly higher for H$\_2$THPP on Ag(110) compared to Au(111) due to the lower desorption rate.\cite{Bai2010}
Subsequently, the influence of the Ag(110) substrate on the molecules is not dominant in the spectra and effects due to heating are more evident. In Fig.~\ref{fig:VB}b, the most interesting region of the highest occupied molecular orbitals before and after annealing of the sample is shown. After the heating step to \SI{200}{\celsius} (green curve) the valence band near $E\_F$ changed slightly (green arrows), similar to observations for H$\_2$TPP\cite{Santo2012}.
The peak centered at \SI{1.6}{eV} corresponds to the convolution of the highest occupied molecular orbital (HOMO) and the HOMO$-$1 (supplemental information, Fig. S3). When the molecular film was annealed to a maximum temperature of \SI{300}{\celsius} ($\approx$ 3 ML remaining), the peak from the highest occupied molecular orbitals broadened significantly and shifted closer to the Fermi energy (blue arrows in Fig.~\hyperref[fig:VB]{\ref*{fig:VB}(b)}). This change of the valence band after \SI{300}{\celsius} treatment is most likely due to the above discussed reaction of C$-$H carbon atoms at the pyrrole subunits and the consequent formation of covalent bonds with adjacent, also deprotonated H$\_2$THPP molecules or with surface atoms. The dissociation of hydroxyl groups on Ag(110) plays likely only a minor role due to the small contribution of the oxygen atoms to the highest molecular orbitals. The change of the amount of H$\_2$THPP relative to H$\_4$THPP molecules might be the reason for the differences in the valence band after annealing to \SI{200}{\celsius}. For changes after heating to \SI{300}{\celsius}, the observed increase of the amount of H$\_4$THPP \textit{via} hydrogen transfer to N cannot be the main reason because the measured valence band clearly differs from the one of the molecular film prior to the first annealing step, where also a higher amount of $-$NH$-$ compared to $-$N= nitrogen was measured.

\section{Conclusions}

N 1s core level spectra showed that the amount of pyrrolic and iminic nitrogen was not the same as expected for H$\_2$THPP, which indicates that H$\_2$THPP and H$\_4$THPP molecules coexisted in the layer. The two additional central hydrogen atoms can be removed by heating due to their relatively low N$-$H dissociation energy. But, on the more reactive Ag(110) surface at higher annealing temperatures, the hydroxyl groups can dissociate and hydrogen atoms subsequently diffuse to the center of the molecule and bond to the nitrogen atoms, which results in the formation of zwitterionic H$\_4$THPP. For heating to and above \SI{300}{\celsius}, partial C$-$H bond cleavage linked with the formation of covalent bonds to other molecules or the surface was observed, which affects the whole electronic structure of the molecule. This, as well as the catalytic properties of the substrate, should be considered if a molecular monolayer is prepared by multilayer desorption or annealing is applied to achieve large domains of self-assembled adsorbate structures for molecular devices.

\section*{Acknowledgements}

This study has been financially supported by the Deutsche Forschungsgemeinschaft (DFG) through Research Unit FOR 1154 and the Fonds der Chemischen Industrie (FCI).
Photoelectron spectroscopy was performed at the Material Science beamline at the synchrotron Elettra (Trieste, Italy). We thank Martin Vondr$\acute{\mathrm{a}}$\v{c}ek for technical assistance.


\bibliographystyle{model1a-num-names}
\bibliography{THPP_PES}

\end{document}